\documentclass[%
superscriptaddress,
 amsmath,amssymb,
 aps,
 prl,
 twocolumn,
]{revtex4-1}

\usepackage{graphicx}
\usepackage{dcolumn}
\usepackage{bm}
\usepackage{natbib}

\usepackage{color}

\begin{document}

\title{Can hail and rain nucleate cloud droplets?}

\author{Prasanth Prabhakaran}
\affiliation{Max Planck Institute for Dynamics and Self-Organization, 37077 G\"ottingen, Germany}
\affiliation{Institute for Nonlinear Dynamics, University of G\"ottingen, 37073 G\"ottingen, Germany}
\author{Stephan Weiss}
\affiliation{Max Planck Institute for Dynamics and Self-Organization, 37077 G\"ottingen, Germany}

\author{Alain Pumir}
\affiliation{Max Planck Institute for Dynamics and Self-Organization, 37077 G\"ottingen, Germany}
\affiliation{Laboratoire de Physique, Ecole Normale Sup\'erieure de Lyon, Universit\'e de Lyon 1 and Centre National de la Recherche Scientifique, 69007 Lyon, France}

\author{Alexei Krekhov}
\affiliation{Max Planck Institute for Dynamics and Self-Organization, 37077 G\"ottingen, Germany}

\author{Eberhard Bodenschatz}
\affiliation{Max Planck Institute for Dynamics and Self-Organization, 37077 G\"ottingen, Germany}
\affiliation{Institute for Nonlinear Dynamics, University of G\"ottingen, 37073 G\"ottingen, Germany}
\affiliation{Laboratory of Atomic and Solid-State Physics and Sibley School of Mechanical and Aerospace Engineering, Cornell University, Ithaca, New York 14853, USA}

\date{\today}

\begin{abstract}
We present results from moist convection in a mixture of pressurized sulfur hexafluoride (liquid and vapor), and helium (gas) to model the wet and dry components of the earth's atmosphere. To allow for homogeneous nucleation, we operate the experiment close to critical conditions. We report  on the nucleation of microdroplets in the wake of large cold liquid drops falling through the supersaturated atmosphere and show that the  homogeneous nucleation is caused by isobaric cooling of the saturated sulfur hexaflouride vapor. Our results carry over to atmospheric clouds: falling hail and cold rain drops may enhance the heterogeneous nucleation of microdroplets in their wake under supersaturated atmospheric conditions. We also observed that under appropriate conditions settling microdroplets form a rather stable horizontal cloud layer, which separates regions of super- and sub critical saturation.

\end{abstract}

\maketitle{}

%
A key process in cloud dynamics is nucleation, {\em i.e.,} the formation of condensation nuclei under supersaturated conditions that eventually grow to form micrometer size cloud droplets~\cite{pruppacher2010cloud}. It is well known that the cloud dynamics and the formation of precipitation size droplets are strongly influenced by the concentration and the properties of the aerosol particles~\cite{pruppacher2010cloud}. Inspired by laboratory experiments~\cite{schaefer1946,vonnegut1947nucleation}, several field experiments were conducted to enhance precipitation in clouds. In two seminal studies \cite{langmuir1948, KrausSquires:1947} dry ice was dropped on top of developing cumulus clouds, which in most cases triggered explosive cloud growth with significant rainfall in its neighborhood. 
Since then several investigations have been carried out to understand the dynamics associated with nucleation in clouds with and without seeding~\cite{pruppacher2010cloud,Bruintjes_1999,mason1982personal,kopp1983numerical}.

\begin{figure*}
\includegraphics[width=1.0\textwidth]{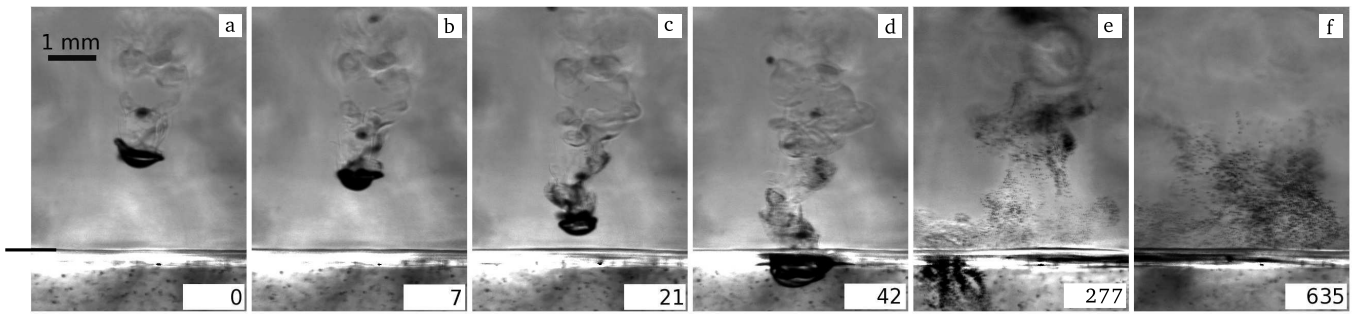}
\caption{\label{fig:wide} 
Contrast enhanced image sequence of an SF$_6$ drop falling through the gaseous SF$_6$-He layer. 
Shown is the lower part of the cell. 
The black mark on the left indicates the position of the liquid-vapor interface located at about $6$\,mm from the bottom plate. 
The time stamp (in ms) for each of these figures is indicated at the bottom right corner.
}
\end{figure*}

In this letter, we report experimental findings on nucleation in a multiphase convection system consisting of sulfur hexafluoride (SF$_6$) and helium (He). 
This system aims to mimic atmospheric conditions with SF$_6$ existing in both liquid and vapor phases, thus acting as the moist component. Helium is added to mimic the dry component in the earth's atmosphere. The advantage of using SF$_6$ is that a relatively small supersaturation is required to trigger homogeneous nucleation \cite{Ye_2000}.
The nucleation of SF$_6$ microdroplets in the wake of cold SF$_6$ drops falling from the top plate (Fig.~\ref{fig:wide}) is our main finding. 
We argue that the induced isobaric cooling in the wake increases locally the saturation ratio, and therefore triggers homogeneous nucleation.
%
We show that a similar mechanism carries over to atmospheric clouds, where falling ice particles (hail) and/or large, cold rain drops can enhance the heterogeneous nucleation rate of droplets. Furthermore, in the experiment, under appropriate conditions, we observe the nucleated SF$_6$ droplets forming a rather stable horizontal cloud layer, separating regions of sub- and super critical saturation, just as in an atmospheric cloud layer.

%

\begin{figure}[t]
\includegraphics[width=\columnwidth]{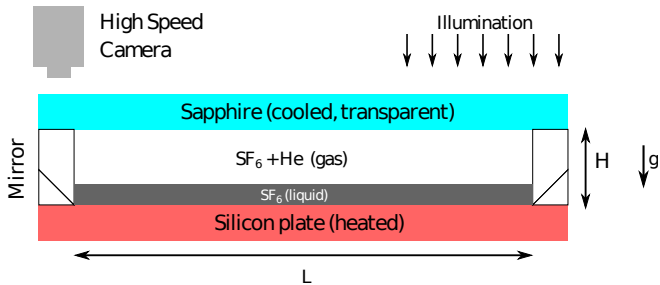}
\caption{\label{fig:epsart} 
Schematic of the experimental setup. 
Mirrors are embedded into the side wall on all four sides of the square cell. 
For clarity only two mirrors are shown.
}
\end{figure}

%
The experiments were performed in a high-pressure convection apparatus that has previously been used to study
pattern formation close to the onset of convection \cite{BBMTHCA96,Plapp1998,WSB14}.
The main part of the apparatus is the convection cell (Fig.~\ref{fig:epsart}) that consists of two
horizontal plates, H=(22.6$\pm0.5$)\,mm apart from each other. The side walls of the cell were made of acrylic
with a square cross section of side length L=(61.65$\pm0.01$)\,mm.
The top plate, a $9.5$\,mm thick monocrystalline sapphire providing optical access, was kept horizontal and was cooled by circulating water on its top surface. 
The bottom plate was a $9.5$\,mm thick monocrystalline silicon disc that was heated with an ohmic film heater at its bottom side. 
The top and bottom plate temperatures were regulated to $\pm$10\,mK of the set temperature. 
In order to provide visual access from the side two sets of mirrors were embedded into the acrylic side walls at 45 degrees. 
One set of mirrors provides optical access to the top half and the other set to the bottom half of the cell (Fig.~\ref{fig:epsart}). Image acquisition was done at 140 fps at a resolution of 2048 x 2048~\cite{camera}.

The bottom plate was heated to a temperature T$_b$, while the top plate was maintained at T$_t$ $<$
T$_b$. The conditions were such that a layer of liquid SF$_6$ formed at the bottom of the cell. 
At the liquid-gas interface, SF$_6$ evaporated, rose and condensed underneath the top plate creating a thin liquid film,
which continuously underwent a Rayleigh-Taylor like instability. 
As a consequence, drops of SF$_6$ dripped and fell through the gas layer into the liquid SF$_6$ pool at the bottom. 

Figure~\ref{fig:wide} shows a deformed drop with a lateral diameter about 1 mm, accompanied by a smaller drop falling through the gaseous SF$_6$-He atmosphere. 
Since these drops originated from the liquid layer at the top plate, their temperature is close to T$_{t}$. 
%
%
Local inhomogeneities in the gas temperature, hence in the refractive index, are visible in the wake of
the cold drop by shadowgraphy (Fig.~\ref{fig:wide}(a-d)).
In the near wake of the deformed drop, the shed vortices mix the cold gas from the boundary layer with the warmer ambient gas as shown in Fig.~\ref{fig:wide}(a-d), thus locally altering the saturation ratio and temperature. 
Particularly in Fig.~\ref{fig:wide}c, the enhanced contrast in the near wake is due to the nucleation of microdroplets which become visible in Fig.~\ref{fig:wide}d. Please note that the large drop enters the liquid pool without a visible splash.
%
These microdroplets continue to grow in size by condensation of SF$_6$ vapor from their supersaturated neighborhood till they fall into the liquid pool (Fig.~\ref{fig:wide} (e,f) and movie in \cite{suppl}). 
Note that for fixed experimental conditions the number of nucleated microdroplets vary strongly for different falling drops (see movie in \cite{suppl}). 
This points to a nonuniform distribution of SF$_6$ vapor due to the turbulent convection in the gaseous layer.
%

%
In the experiment shown in Fig.~\ref{fig:wide}, the temperature at the top and bottom plates were T$_{t} =40.00\,^\circ$C and T$_{b} =44.00 \,^\circ$C, and the pressure was $p =(46.9\pm0.1$)\,bar. 
Based on Dalton's law, 
we estimate the mole fraction of He inside the gaseous layer to be $x_{He} \approx 26$\%. 
We found that the nucleation of microdroplets in the wake of a falling drop was observed when T$_b$ was sufficiently close to the critical temperature of SF$_6$ (45.57\,$^\circ$C) at a fixed T$_b$ - T$_t$. 
As we show below, this can be attributed to the lowering of the critical supersaturation required for nucleation as the critical temperature is approached. 
%

%
%
Classical nucleation theory~\cite{Feder_1966,Zeldovich_1943} provides an estimate for the rate of formation of liquid phase critical droplets ("embryos"), $J$, as a function of the saturation ratio, $S$.
By convention \cite{pruppacher2010cloud}, the detectable rate of nucleation is taken to be $J_c = 1$ cm$^{-3}$s$^{-1}$. This leads to a definition of a critical saturation ratio $S_c$. 
The corresponding critical size $r_c$ that needs to be exceeded for a sustained droplet growth is calculated using the Kelvin's equation ~\cite{suppl}. 

For the mean temperature $T_m =42 \,^\circ$C, using SF$_6$ parameters \cite{Guder_2009,RefProp,details1}, we find $S_c =1.000815$, $r_c =21.2$~nm and the time to establish the steady-state nucleation rate $\tau =1.3$~$\mu$s (see \cite{suppl}). 
%
%
Using Maxwell's model (diffusion limited growth), we estimate the time for growth by condensation from an initial radius of $r_c$ to $r =10$~$\mu$m to $41.5$~ms ~\cite{suppl}. This time agrees remarkably well with
the time difference between Fig.~\ref{fig:wide}(a) and Fig.~\ref{fig:wide}(d), i.e., the time interval between the cooling of the
ambient gas at a
certain location and the first appearance of microdroplets in its neighborhood.
%

%
In the experiment, the SF$_6$ vapor close to a falling cold drop is cooled from an initial
temperature $T$ to $T - \Delta T$ due to diffusive and convective heat transport from the drop's surface.
As a consequence, the saturation ratio becomes $S =p_{v}(T)/p_{s}(T - \Delta T)$, where $\Delta T$ represents the temperature difference between the ambient gas and the wake. 
For $T_m =42\,^\circ$C of saturated SF$_6$ vapor, i.e. $p_{v}(T) = p_{s}(T)$, a $\Delta T =0.04$~K is sufficient to reach $S =S_c$, in comparison to a $\Delta T=0.57$~K at $T_m=30\,^\circ$C.
Note that the cooling in the wake to $\Delta T > 0.04$~K is attainable, given that the temperature of the cold falling drop was initially $2$~K below the mean temperature in the cell. 
To show this, we estimate the cooling, i.e., the temperature difference 
$\Delta$ between the ambient gas and a drop of diameter $d$, falling 
at its terminal velocity $U_t$, in a gas layer of temperature $T_a(z)$ 
decreasing linearly with height. 
We find:
%
%
\begin{eqnarray}
\label{eq:Delta_t}
\Delta = \Delta_0e^{-A t} + \frac{\beta U_t}{A} (1 - e^{-A t}) 
\quad \textrm{with} \quad 
A = \frac{6 \lambda \textrm{Nu}}{\rho_{l} d^2 c_{p, l}} \,
\end{eqnarray}
where $\Delta_0 =\Delta(t=0)$, $\beta =d T_a/d z$, $\textrm{Nu}$ is the Nusselt number (ratio between convective and conductive heat transfer), $\lambda$ is the thermal conductivity of the ambient gas, $\rho_{l}$, $c_{p, l}$ are the density, and the specific heat of the liquid respectively (see \cite{suppl} for additional details).

In Fig.~\ref{fig:wide}, the diameter of the cold SF$_6$ drops detaching from the top plate $d \approx 0.5$~mm. 
From the recorded images, we found that the terminal velocity of the drop U$_t$ $\approx$ 7 cm/s was reached after about a $2$~mm fall from the top plate. 
Drops reach the liquid layer above the bottom plate in about $0.2$~s.
Using the material parameters of SF$_6$ at $T_m =42\,^\circ$C \cite{details1} and an empirical relation between the drop Reynolds number (Re$ =U_t d/\nu$), and Nu~\cite{whitaker1972forced}, we find $\textrm{Re} \approx 600$ and $\textrm{Nu} \approx 26$, and thus from Eq.~(\ref{eq:Delta_t}), $1/A \approx 0.18$~s.
Let us assume that when the drop attains its terminal velocity ($t = 0$), it has the same temperature as its ambient, i.e., $\Delta_0=0$.
To account for convective mixing in the gas layer we choose $\beta =0.5$~K/cm, which is four times smaller than the applied temperature gradient of $2$~K/cm across the gas layer.
Equation~(\ref{eq:Delta_t}) predicts that $\Delta = 0.1$~K and $0.2$~K  at $t \approx 30$~ms and $60$~ms, respectively, which is well before the drop enters the liquid pool.

The cooling $\Delta T$ in the wake of the drop is determined by the heat transfer rate from the ambient gas to
the cold drop and as such is a function of Re of the falling drop and the
streamwise distance from the drop's surface.
Simulations at $\textrm{Re} >300$ \cite{de2014large} of the instantaneous temperature distribution in the near wake show that the separated shear layers retain up to $20$\% of $\Delta$ till about $2$ droplet diameters downstream. 
The settling cold drop, with $\Delta =0.2$~K, therefore induces isobaric cooling of the ambient wake by $\Delta T \approx 0.04$K at $t=60$~ms which is sufficient to trigger homogeneous nucleation at $T_m = 42\,^\circ$C. 
The cooling in the near wake is enhanced with further fall of the drop.

Please note that this is a simple estimate. In fact, the $\Delta T$ required to trigger homogeneous nucleation also depends on the distribution of SF$_6$ vapor in the boundary layer and the wake. Moreover, additional complexities arise due to the mixing of parcels of different temperature and vapor content. 
We here assume that the SF$_6$ vapor content is constant and saturated at $T_m$. As a consequence, the level of supersaturation estimated by isobaric cooling with constant SF$_6$ vapor content gives an upper bound.

\begin{figure}[b]
\includegraphics[width=\columnwidth]{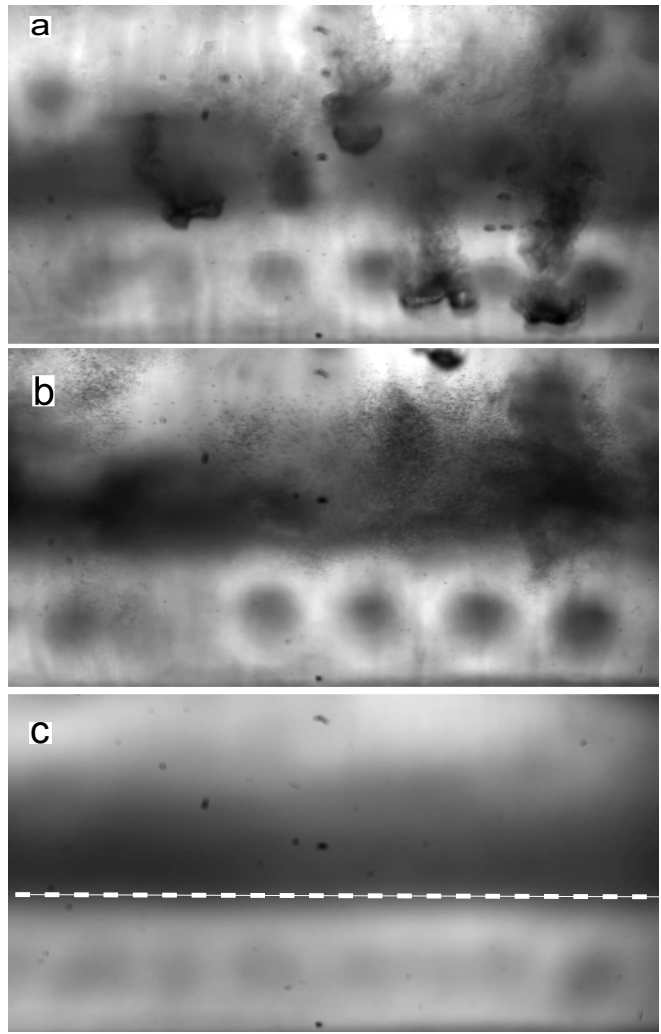}
\caption{\label{fig:layer} 
Cloud layer: Formation of a cloud layer in the lower part of the cell in the absence of a liquid layer at
the bottom plate. 
T$_b = 45.00\,^\circ$C, T$_t = 36.00\,^\circ$C, $p = 44.9$\,bar. 
The dark circular patches in the lower part of the image are due to the drops attached to the top plate. 
(a) Drops from the top plate falling through the gas layer. 
(b) Microdroplets nucleated in the wake and suspended in the gas. Image taken 760\,ms
after (a).
(c) Time averaged image. 
The white dashed line represents the mean position of the base of the cloud layer and thus
vertically separates areas of super- (above) and sub critical saturation (below).
}
\end{figure}

%
%
If the bottom plate is covered by a liquid layer, the SF$_6$ vapor in the gas is on average saturated or slightly supersaturated due to the continuous supply of vapor from the liquid pool below. 
As a consequence, any sufficiently large microdroplet would continue to grow till it reaches the liquid layer. 
The saturation ratio $S$, of SF$_6$ vapor inside the cell can be lowered by eliminating the liquid layer above the bottom plate while keeping all the other parameters fixed, thus cutting off the vapor supply. %
In such cases, $S \ge S_c$ in the upper, colder part of the cell, while $S < S_c$ in the warmer lower part. 
Figure~\ref{fig:layer} displays the observations in the lower part of the cell in the absence of a liquid layer on the bottom. 
The microdroplets were nucleated in the wake of drops falling from the top plate, similar to that in
Fig.~\ref{fig:wide}, except that sustained nucleation in the wake was observed only above a certain height (see Fig.~\ref{fig:layer}a, b and movie in \cite{suppl}). 
This height marks the horizontal interface where $S \approx S_c$. 
In the region above this interface, the nucleated microdroplets grow in size and below the interface they evaporate (see movie in \cite{suppl}). 
Figure~\ref{fig:layer}b shows a layer of microdroplets suspended in the gas layer. 
The dark band in Fig.~\ref{fig:layer}c represents the time averaged cloud layer and it suggests that the layer has a well defined base, similar to the clouds in the earth's atmosphere.
The mean position (averaged over 1550 frames, $\approx 11$ seconds) of this base is marked in Fig.~\ref{fig:layer}c. 
Below this interface no sustained nucleation was observed. 
%

%
Let us now compare the conditions in the earth's atmosphere with those in the experiment
and check under which conditions the present observations can be carried over.
Due to the much larger surface tension of water in comparison to SF$_6$, the saturation ratio required to achieve $J_c$ for water vapor is much higher. 
For example, for homogeneous nucleation of liquid water in the moist air for the temperature range $-30\,^\circ$C~$ \le T_a \le +30\,^\circ$C the critical saturation ratio varies between $7$ and $3$. 
However, the supersaturation that develops in natural clouds rarely exceeds a few percent \cite{pruppacher2010cloud}. 
Consequently, water droplets do not form by homogeneous nucleation but rather by heterogeneous nucleation on atmospheric aerosol particles \cite{pruppacher2010cloud}. 
The number of particles capable of growing, denoted as cloud condensation nuclei (CCN) grows with supersaturation, $s = S - 1$, typically with a power law dependence on $s$~\cite{pruppacher2010cloud}.
Consider moist cloudy air at temperature $T$ and $S \approx 1$ that contains a certain amount of the CCN. 
When a large cold water drop (or hail particle), falls through a
warmer ambient air, it will cause isobaric cooling of the air in its wake, similar to the drops in
our experiment.
This results in locally enhanced supersaturation that can create more CCN, and an increase in the number of microdroplets. 
For example, to attain a supersaturation $s =2$\% by isobaric cooling starting from the saturated vapor, a temperature drop of $0.21$~K$\le \Delta T \le 0.34$~K (varies almost linearly) for the temperature range $-30^\circ$C~$ \le T_a \le +30^\circ$C is needed.
In a cloud, the ambient temperature surrounding a falling rain drop increases as the drop approaches the cloud base. 
We assume that the temperature variation within a cloud is linear with a typical moist adiabatic lapse rate $\approx 0.005$~K/m~\cite{bohren_1998}. 
Then for a raindrop at its terminal velocity $U_t$, the steady state temperature difference between the ambient air and the drop, $\Delta= \Delta_{\infty}= {\beta U_t}/{A}$, is calculated using Eq.~(\ref{eq:Delta_t}). 
For a drop of diameter $1$\,mm ($5$\,mm), one finds $\Delta_{\infty}$ is about $0.07$\,K ($1.3$\,K) (see \cite{suppl}). 
The $\Delta_\infty$ for the $1$\,mm drop is too small to cause significant
supersaturation in its wake. However, for a $5$\,mm drop, the resulting cooling in the wake, $\Delta T = 0.2~\Delta_{\infty}=0.26$~K is sufficient to attain about $2$\% supersaturation and thus leads to enhanced nucleation in its wake. 
%

%

%
In the case of a falling ice particle, as the ambient temperature increases above $0\,^\circ$C, ice particles begin to melt, a process that extracts heat from its surrounding, thus resulting in a supersaturation in the wake that is higher than that of a rain drop of the same diameter.
%
%
It is known that the temperature inside a particle composed of a mixture of liquid water and ice is nearly homogeneous due to the shear enhanced mixing inside the particle \cite{rasmussen1984wind}. 
As a consequence the temperature of the ice particle would not increase until it is completely melted. 
The heat transfer rate is also known to depend on the shape of the ice particle \cite{rasmussen1984wind}. 
Let us assume a spherical particle of uniform density. To account for liquid condensation on the
surface, we further assume a 25\% larger heat transfer rate between the particle and its
surrounding than for the liquid drop \cite{mason1956melting}.
Based on these considerations an ice particle of $1$\,mm diameter ($\textrm{Re} \approx 230$) would travel $\approx 450$\,m and a particle of diameter $5$\,mm ($\textrm{Re} \approx 3600$) would travel $\approx 2000$\,m before it is completely melted. 
The corresponding maximum $\Delta_{\infty}$ for these particles are about $2.2$\,K and $10$\,K, respectively (see \cite{suppl} for details). 
As a consequence, the local supersaturation would be around $2 - 10$\% in the near wake of the particle taking into account $\Delta T = 0.2~\Delta_{\infty}$. 
This enhanced supersaturation would activate more nuclei and hence increase the concentration of microdroplets in the warmer part of the cloud. 
Note, that for an aerosol particle with radius $\le 1$~$\mu$m, the time required to activate the CCN is smaller than the Kolmogorov time scale in the clouds \cite{pruppacher2010cloud}. 
The estimates here are based on idealized conditions. 
In a (convective) cloud, the dynamics is more complicated due to the presence of updrafts, variable lapse rates due to non-uniform latent heat release from condensation/glaciation, and inhomogeneous mixing due to entrainment of ambient dry air into the cloud \cite{lehmann2009homogeneous}. 
Nevertheless, the results from our model system and the analysis presented in this letter suggest that in clouds, the cooling induced by a falling hail particle can indeed lead to the nucleation of droplets. 
This effect may play an important role, as the droplets produced by this mechanism may either collide and aggregate with other settling hail or rain drops, or be entrained into an updraft, to further reinforce the production of hail or large rain drops. 
Moreover, the additional latent heat released due to the nucleation of new droplets can feed energy to the existing updraft. 

The results presented in this letter revealed an unexpected mechanism of nucleation and growth of microdroplets in nonequlibirum conditions, such as those in the atmosphere. Our estimates predict that the enhanced nucleation of small droplets by a cold falling drop or ice particle, former clearly observed in the experiments, should also play a role in clouds. It is worth noting that the ideas developed here could be potentially extended to the nucleation of small ice crystals in the wake of large hail particles or graupels. Testing the ideas presented here will require additional experiments under atmospheric conditions.


\begin{thebibliography}{25}%
\makeatletter
\providecommand \@ifxundefined [1]{%
 \@ifx{#1\undefined}
}%
\providecommand \@ifnum [1]{%
 \ifnum #1\expandafter \@firstoftwo
 \else \expandafter \@secondoftwo
 \fi
}%
\providecommand \@ifx [1]{%
 \ifx #1\expandafter \@firstoftwo
 \else \expandafter \@secondoftwo
 \fi
}%
\providecommand \natexlab [1]{#1}%
\providecommand \enquote  [1]{``#1''}%
\providecommand \bibnamefont  [1]{#1}%
\providecommand \bibfnamefont [1]{#1}%
\providecommand \citenamefont [1]{#1}%
\providecommand \href@noop [0]{\@secondoftwo}%
\providecommand \href [0]{\begingroup \@sanitize@url \@href}%
\providecommand \@href[1]{\@@startlink{#1}\@@href}%
\providecommand \@@href[1]{\endgroup#1\@@endlink}%
\providecommand \@sanitize@url [0]{\catcode `\\12\catcode `\$12\catcode
  `\&12\catcode `\#12\catcode `\^12\catcode `\_12\catcode `\%12\relax}%
\providecommand \@@startlink[1]{}%
\providecommand \@@endlink[0]{}%
\providecommand \url  [0]{\begingroup\@sanitize@url \@url }%
\providecommand \@url [1]{\endgroup\@href {#1}{\urlprefix }}%
\providecommand \urlprefix  [0]{URL }%
\providecommand \Eprint [0]{\href }%
\providecommand \doibase [0]{http://dx.doi.org/}%
\providecommand \selectlanguage [0]{\@gobble}%
\providecommand \bibinfo  [0]{\@secondoftwo}%
\providecommand \bibfield  [0]{\@secondoftwo}%
\providecommand \translation [1]{[#1]}%
\providecommand \BibitemOpen [0]{}%
\providecommand \bibitemStop [0]{}%
\providecommand \bibitemNoStop [0]{.\EOS\space}%
\providecommand \EOS [0]{\spacefactor3000\relax}%
\providecommand \BibitemShut  [1]{\csname bibitem#1\endcsname}%
\let\auto@bib@innerbib\@empty
\bibitem [{\citenamefont {Pruppacher}\ and\ \citenamefont
  {Klett}(2010)}]{pruppacher2010cloud}%
  \BibitemOpen
  \bibfield  {author} {\bibinfo {author} {\bibfnamefont {H.}~\bibnamefont
  {Pruppacher}}\ and\ \bibinfo {author} {\bibfnamefont {J.}~\bibnamefont
  {Klett}},\ }\href@noop {} {\emph {\bibinfo {title} {Microphysics of Clouds
  and Precipitation}}}\ (\bibinfo  {publisher} {Springer},\ \bibinfo {year}
  {2010})\BibitemShut {NoStop}%
\bibitem [{\citenamefont {Schaefer}(1946)}]{schaefer1946}%
  \BibitemOpen
  \bibfield  {author} {\bibinfo {author} {\bibfnamefont {V.~J.}\ \bibnamefont
  {Schaefer}},\ }\href@noop {} {\bibfield  {journal} {\bibinfo  {journal}
  {Science}\ }\textbf {\bibinfo {volume} {104}},\ \bibinfo {pages} {457}
  (\bibinfo {year} {1946})}\BibitemShut {NoStop}%
\bibitem [{\citenamefont {Vonnegut}(1947)}]{vonnegut1947nucleation}%
  \BibitemOpen
  \bibfield  {author} {\bibinfo {author} {\bibfnamefont {B.}~\bibnamefont
  {Vonnegut}},\ }\href@noop {} {\bibfield  {journal} {\bibinfo  {journal} {J.
  Appl. Phys.}\ }\textbf {\bibinfo {volume} {18}},\ \bibinfo {pages} {593}
  (\bibinfo {year} {1947})}\BibitemShut {NoStop}%
\bibitem [{\citenamefont {Langmuir}(1948)}]{langmuir1948}%
  \BibitemOpen
  \bibfield  {author} {\bibinfo {author} {\bibfnamefont {I.}~\bibnamefont
  {Langmuir}},\ }\href@noop {} {\bibfield  {journal} {\bibinfo  {journal} {J.
  Meteorol.}\ }\textbf {\bibinfo {volume} {5}},\ \bibinfo {pages} {175}
  (\bibinfo {year} {1948})}\BibitemShut {NoStop}%
\bibitem [{\citenamefont {Kraus}\ and\ \citenamefont
  {Squires}(1947)}]{KrausSquires:1947}%
  \BibitemOpen
  \bibfield  {author} {\bibinfo {author} {\bibfnamefont {E.~B.}\ \bibnamefont
  {Kraus}}\ and\ \bibinfo {author} {\bibfnamefont {P.}~\bibnamefont
  {Squires}},\ }\href@noop {} {\bibfield  {journal} {\bibinfo  {journal}
  {Nature}\ }\textbf {\bibinfo {volume} {159}},\ \bibinfo {pages} {489}
  (\bibinfo {year} {1947})}\BibitemShut {NoStop}%
\bibitem [{\citenamefont {Bruintjes}(1999)}]{Bruintjes_1999}%
  \BibitemOpen
  \bibfield  {author} {\bibinfo {author} {\bibfnamefont {R.~T.}\ \bibnamefont
  {Bruintjes}},\ }\href@noop {} {\bibfield  {journal} {\bibinfo  {journal}
  {Bull. Amer. Meteor. Soc.}\ }\textbf {\bibinfo {volume} {80}},\ \bibinfo
  {pages} {805} (\bibinfo {year} {1999})}\BibitemShut {NoStop}%
\bibitem [{\citenamefont {Mason}(1982)}]{mason1982personal}%
  \BibitemOpen
  \bibfield  {author} {\bibinfo {author} {\bibfnamefont {B.}~\bibnamefont
  {Mason}},\ }\href@noop {} {\bibfield  {journal} {\bibinfo  {journal}
  {Contemp. Phys.}\ }\textbf {\bibinfo {volume} {23}},\ \bibinfo {pages} {311}
  (\bibinfo {year} {1982})}\BibitemShut {NoStop}%
\bibitem [{\citenamefont {Kopp}\ \emph {et~al.}(1983)\citenamefont {Kopp},
  \citenamefont {Orville}, \citenamefont {Farley},\ and\ \citenamefont
  {Hirsch}}]{kopp1983numerical}%
  \BibitemOpen
  \bibfield  {author} {\bibinfo {author} {\bibfnamefont {F.~J.}\ \bibnamefont
  {Kopp}}, \bibinfo {author} {\bibfnamefont {H.~D.}\ \bibnamefont {Orville}},
  \bibinfo {author} {\bibfnamefont {R.~D.}\ \bibnamefont {Farley}}, \ and\
  \bibinfo {author} {\bibfnamefont {J.~H.}\ \bibnamefont {Hirsch}},\
  }\href@noop {} {\bibfield  {journal} {\bibinfo  {journal} {J. Appl.
  Meteorol.}\ }\textbf {\bibinfo {volume} {22}},\ \bibinfo {pages} {1542}
  (\bibinfo {year} {1983})}\BibitemShut {NoStop}%
\bibitem [{\citenamefont {Ye}\ \emph {et~al.}(2000)\citenamefont {Ye},
  \citenamefont {Bertelsmann}, \citenamefont {Heist}, \citenamefont {Hale},\
  and\ \citenamefont {Kulmala}}]{Ye_2000}%
  \BibitemOpen
  \bibfield  {author} {\bibinfo {author} {\bibfnamefont {P.}~\bibnamefont
  {Ye}}, \bibinfo {author} {\bibfnamefont {A.}~\bibnamefont {Bertelsmann}},
  \bibinfo {author} {\bibfnamefont {R.~H.}\ \bibnamefont {Heist}}, \bibinfo
  {author} {\bibfnamefont {B.~N.}\ \bibnamefont {Hale}}, \ and\ \bibinfo
  {author} {\bibfnamefont {M.}~\bibnamefont {Kulmala}},\ }in\ \href@noop {}
  {\emph {\bibinfo {booktitle} {AIP Conference Proceedings}}},\ Vol.\ \bibinfo
  {volume} {534}\ (\bibinfo {organization} {AIP},\ \bibinfo {year} {2000})\
  pp.\ \bibinfo {pages} {19--22}\BibitemShut {NoStop}%
\bibitem [{\citenamefont {de~Bruyn}\ \emph {et~al.}(1996)\citenamefont
  {de~Bruyn}, \citenamefont {{E.~Bodenschatz, S.~W.~Morris, S.~Trainoff,
  Y.~Hu}}, \citenamefont {Cannell},\ and\ \citenamefont {Ahlers}}]{BBMTHCA96}%
  \BibitemOpen
  \bibfield  {author} {\bibinfo {author} {\bibfnamefont {J.~R.}\ \bibnamefont
  {de~Bruyn}}, \bibinfo {author} {\bibnamefont {{E.~Bodenschatz, S.~W.~Morris,
  S.~Trainoff, Y.~Hu}}}, \bibinfo {author} {\bibfnamefont {D.~S.}\ \bibnamefont
  {Cannell}}, \ and\ \bibinfo {author} {\bibfnamefont {G.}~\bibnamefont
  {Ahlers}},\ }\href@noop {} {\bibfield  {journal} {\bibinfo  {journal} {Rev.
  Sci. Instrum.}\ }\textbf {\bibinfo {volume} {67}},\ \bibinfo {pages} {2043}
  (\bibinfo {year} {1996})}\BibitemShut {NoStop}%
\bibitem [{\citenamefont {Plapp}\ \emph {et~al.}(1998)\citenamefont {Plapp},
  \citenamefont {Egolf}, \citenamefont {Bodenschatz},\ and\ \citenamefont
  {Pesch}}]{Plapp1998}%
  \BibitemOpen
  \bibfield  {author} {\bibinfo {author} {\bibfnamefont {B.~B.}\ \bibnamefont
  {Plapp}}, \bibinfo {author} {\bibfnamefont {D.~A.}\ \bibnamefont {Egolf}},
  \bibinfo {author} {\bibfnamefont {E.}~\bibnamefont {Bodenschatz}}, \ and\
  \bibinfo {author} {\bibfnamefont {W.}~\bibnamefont {Pesch}},\ }\href
  {\doibase 10.1103/PhysRevLett.81.5334} {\bibfield  {journal} {\bibinfo
  {journal} {Phys. Rev. Lett.}\ }\textbf {\bibinfo {volume} {81}},\ \bibinfo
  {pages} {5334} (\bibinfo {year} {1998})}\BibitemShut {NoStop}%
\bibitem [{\citenamefont {Weiss}\ \emph {et~al.}(2014)\citenamefont {Weiss},
  \citenamefont {Seiden},\ and\ \citenamefont {Bodenschatz}}]{WSB14}%
  \BibitemOpen
  \bibfield  {author} {\bibinfo {author} {\bibfnamefont {S.}~\bibnamefont
  {Weiss}}, \bibinfo {author} {\bibfnamefont {G.}~\bibnamefont {Seiden}}, \
  and\ \bibinfo {author} {\bibfnamefont {E.}~\bibnamefont {Bodenschatz}},\
  }\href {\doibase 10.1017/jfm.2014.456} {\bibfield  {journal} {\bibinfo
  {journal} {J. Fluid Mech.}\ }\textbf {\bibinfo {volume} {756}},\ \bibinfo
  {pages} {293} (\bibinfo {year} {2014})}\BibitemShut {NoStop}%
\bibitem [{cam()}]{camera}%
  \BibitemOpen
  \href@noop {} {}\bibinfo {howpublished} {Phantom 65 Gold Camera, Vision
  Research.}\BibitemShut {Stop}%
\bibitem [{sup()}]{suppl}%
  \BibitemOpen
  \href@noop {} {}\bibinfo {howpublished} {See Supplemental Material for
  additional details and movies, which includes Ref~\cite{clift1971motion}}\BibitemShut {NoStop}%
\bibitem[{\citenamefont{Clift and Gauvin}(1971)}]{clift1971motion}
  \bibinfo{author}{\bibfnamefont{R.}~\bibnamefont{Clift}} \bibnamefont{and}
  \bibinfo{author}{\bibfnamefont{W.}~\bibnamefont{Gauvin}},
  \bibinfo{journal}{The Canadian Journal of Chemical Engineering}
  \textbf{\bibinfo{volume}{49}}, \bibinfo{pages}{439} (\bibinfo{year}{1971}).
\bibitem [{\citenamefont {Feder}\ \emph {et~al.}(1966)\citenamefont {Feder},
  \citenamefont {Russell}, \citenamefont {Lothe},\ and\ \citenamefont
  {Pound}}]{Feder_1966}%
  \BibitemOpen
  \bibfield  {author} {\bibinfo {author} {\bibfnamefont {J.}~\bibnamefont
  {Feder}}, \bibinfo {author} {\bibfnamefont {K.~C.}\ \bibnamefont {Russell}},
  \bibinfo {author} {\bibfnamefont {J.}~\bibnamefont {Lothe}}, \ and\ \bibinfo
  {author} {\bibfnamefont {G.~M.}\ \bibnamefont {Pound}},\ }\href@noop {}
  {\bibfield  {journal} {\bibinfo  {journal} {Adv. Phys.}\ }\textbf {\bibinfo
  {volume} {15}},\ \bibinfo {pages} {111} (\bibinfo {year} {1966})}\BibitemShut
  {NoStop}%
\bibitem [{\citenamefont {Zeldovich}(1943)}]{Zeldovich_1943}%
  \BibitemOpen
  \bibfield  {author} {\bibinfo {author} {\bibfnamefont {J.~B.}\ \bibnamefont
  {Zeldovich}},\ }\href@noop {} {\bibfield  {journal} {\bibinfo  {journal}
  {Acta Physicochimica URSS}\ }\textbf {\bibinfo {volume} {18}},\ \bibinfo
  {pages} {1} (\bibinfo {year} {1943})}\BibitemShut {NoStop}%
\bibitem [{\citenamefont {Guder}\ and\ \citenamefont
  {Wagner}(2009)}]{Guder_2009}%
  \BibitemOpen
  \bibfield  {author} {\bibinfo {author} {\bibfnamefont {C.}~\bibnamefont
  {Guder}}\ and\ \bibinfo {author} {\bibfnamefont {W.}~\bibnamefont {Wagner}},\
  }\href {\doibase 10.1063/1.3037344} {\bibfield  {journal} {\bibinfo
  {journal} {J. Phys. Chem. Ref. Data}\ }\textbf {\bibinfo {volume} {38}},\
  \bibinfo {pages} {33} (\bibinfo {year} {2009})}\BibitemShut {NoStop}%
\bibitem [{\citenamefont {Lemmon}\ \emph {et~al.}(2013)\citenamefont {Lemmon},
  \citenamefont {Huber},\ and\ \citenamefont {McLinden}}]{RefProp}%
  \BibitemOpen
  \bibfield  {author} {\bibinfo {author} {\bibfnamefont {E.~W.}\ \bibnamefont
  {Lemmon}}, \bibinfo {author} {\bibfnamefont {M.~L.}\ \bibnamefont {Huber}}, \
  and\ \bibinfo {author} {\bibfnamefont {M.~O.}\ \bibnamefont {McLinden}},\
  }\href {http://www.nist.gov/srd/refprop} {\emph {\bibinfo {title} {NIST
  Standard Reference Database 23: Reference Fluid Thermodynamic and Transport
  Properties - REFPROP, Version 9.1}}},\ \bibinfo {organization} {{National
  Institute of Standards and Technology, Standard Reference Data Program}},\
  \bibinfo {address} {Gaithersburg} (\bibinfo {year} {2013})\BibitemShut
  {NoStop}%
\bibitem [{det()}]{details1}%
  \BibitemOpen
  \href@noop {} {}\bibinfo {howpublished} {Material parameters of SF$_6$ at $T
  =42\,^\circ$C: $p_{s}=34.715$~bar, $\rho_{l} =1053.7$~kg/m$^3$, $\rho_{v}
  =445.05$~kg/m$^3$, $M =146.0554$~g/mol, $\sigma =0.1629$~mN/m.}\BibitemShut
  {Stop}%
\bibitem [{\citenamefont {Whitaker}(1972)}]{whitaker1972forced}%
  \BibitemOpen
  \bibfield  {author} {\bibinfo {author} {\bibfnamefont {S.}~\bibnamefont
  {Whitaker}},\ }\href@noop {} {\bibfield  {journal} {\bibinfo  {journal}
  {AIChE Journal}\ }\textbf {\bibinfo {volume} {18}},\ \bibinfo {pages} {361}
  (\bibinfo {year} {1972})}\BibitemShut {NoStop}%
\bibitem [{\citenamefont {de~Stadler}\ \emph {et~al.}(2014)\citenamefont
  {de~Stadler}, \citenamefont {Rapaka},\ and\ \citenamefont
  {Sarkar}}]{de2014large}%
  \BibitemOpen
  \bibfield  {author} {\bibinfo {author} {\bibfnamefont {M.~B.}\ \bibnamefont
  {de~Stadler}}, \bibinfo {author} {\bibfnamefont {N.~R.}\ \bibnamefont
  {Rapaka}}, \ and\ \bibinfo {author} {\bibfnamefont {S.}~\bibnamefont
  {Sarkar}},\ }\href@noop {} {\bibfield  {journal} {\bibinfo  {journal} {Int.
  J. Heat Fluid Flow}\ }\textbf {\bibinfo {volume} {49}},\ \bibinfo {pages} {2}
  (\bibinfo {year} {2014})}\BibitemShut {NoStop}%
\bibitem [{\citenamefont {Bohren}\ and\ \citenamefont
  {Albrecht}(1998)}]{bohren_1998}%
  \BibitemOpen
  \bibfield  {author} {\bibinfo {author} {\bibfnamefont {C.~F.}\ \bibnamefont
  {Bohren}}\ and\ \bibinfo {author} {\bibfnamefont {B.~A.}\ \bibnamefont
  {Albrecht}},\ }\href@noop {} {\emph {\bibinfo {title} {Atmospheric
  Thermodynamics}}}\ (\bibinfo  {publisher} {Oxford University Press, New
  York},\ \bibinfo {year} {1998})\BibitemShut {NoStop}%
\bibitem [{\citenamefont {Rasmussen}\ \emph {et~al.}(1984)\citenamefont
  {Rasmussen}, \citenamefont {Levizzani},\ and\ \citenamefont
  {Pruppacher}}]{rasmussen1984wind}%
  \BibitemOpen
  \bibfield  {author} {\bibinfo {author} {\bibfnamefont {R.}~\bibnamefont
  {Rasmussen}}, \bibinfo {author} {\bibfnamefont {V.}~\bibnamefont
  {Levizzani}}, \ and\ \bibinfo {author} {\bibfnamefont {H.}~\bibnamefont
  {Pruppacher}},\ }\href@noop {} {\bibfield  {journal} {\bibinfo  {journal} {J.
  Atmos. Sci}\ }\textbf {\bibinfo {volume} {41}},\ \bibinfo {pages} {381}
  (\bibinfo {year} {1984})}\BibitemShut {NoStop}%
\bibitem [{\citenamefont {Mason}(1956)}]{mason1956melting}%
  \BibitemOpen
  \bibfield  {author} {\bibinfo {author} {\bibfnamefont {B.~J.}\ \bibnamefont
  {Mason}},\ }\href@noop {} {\bibfield  {journal} {\bibinfo  {journal} {Q. J.
  R. Meteorol. Soc.}\ }\textbf {\bibinfo {volume} {82}},\ \bibinfo {pages}
  {209} (\bibinfo {year} {1956})}\BibitemShut {NoStop}%
\bibitem [{\citenamefont {Lehmann}\ \emph {et~al.}(2009)\citenamefont
  {Lehmann}, \citenamefont {Siebert},\ and\ \citenamefont
  {Shaw}}]{lehmann2009homogeneous}%
  \BibitemOpen
  \bibfield  {author} {\bibinfo {author} {\bibfnamefont {K.}~\bibnamefont
  {Lehmann}}, \bibinfo {author} {\bibfnamefont {H.}~\bibnamefont {Siebert}}, \
  and\ \bibinfo {author} {\bibfnamefont {R.~A.}\ \bibnamefont {Shaw}},\
  }\href@noop {} {\bibfield  {journal} {\bibinfo  {journal} {J. Atmos. Sci}\
  }\textbf {\bibinfo {volume} {66}},\ \bibinfo {pages} {3641} (\bibinfo {year}
  {2009})}\BibitemShut {NoStop}%
\end{thebibliography}

%

\appendix
\section{Homogeneous nucleation}

Let us consider the conditions for homogeneous nucleation and growth of microdroplets in SF$_6$ vapor. 
The saturation ratio is defined as the ratio of the vapor pressure and the saturation (equilibrium) vapor pressure at a given temperature, $S =p_{v}/p_{s}$. 
According to classical nucleation theory [16, 17],
 the rate of formation of liquid phase critical droplets (``embryos'') for $S >1$ is given by 
\begin{eqnarray}
\label{eq:NR}
J = \frac{1}{\rho_{l}} \sqrt{\frac{2 M \sigma}{\pi N_{A}}}
\left( \frac{p_{s}}{k T} \right)^2
S \exp\left[ -\frac{16\pi M^2 \sigma^3}{3 (N_{A} \rho_{l} \ln{S})^2 (k T)^3} \right]  \quad
\end{eqnarray}
where $\rho_{l}$ is the density of the liquid phase, $M$ -- its molecular weight, $\sigma$ is the surface tension at the liquid-vapor interface, $N_{A}$ and $k$ are Avogadro's and Boltzmann's constants, respectively.
By convention [1]
, the threshold defining a significant (detectable) rate of homogeneous nucleation is taken to be $J_c =1$~cm$^{-3}$~s$^{-1}$, which in turn leads to the definition of a critical value of the saturation, $S_c$.
In homogeneous nucleation, the critical size of the liquid droplet embryo, $r_c$, required for a sustained droplet growth can be calculated using the Kelvin's equation 
\begin{eqnarray}
\label{eq:rc}
r_c = \frac{2 M \sigma}{k T N_{A} \rho_{l} \ln{S}} \,.
\end{eqnarray}
The time lag to attain steady-state values for the nucleation-rate (see, e.g., [1, 16])  
can be estimated as
\begin{eqnarray}
\label{eq:tauL}
\tau = \frac{\sigma}{\phi k T (\ln{S})^2} \,,
\end{eqnarray}
where $\phi = p_{v}/\sqrt{2\pi (M/N_{A}) k T}$ is the flux of vapor molecules to the droplet embryo surface at the vapor pressure $p_{v}$ and the temperature $T$.
Embryos with $r \ge r_c$ will grow initially by condensation of the molecules from the vapor onto their surface.
According to Maxwell's model (diffusion limited growth), the radius of the droplet changes with time as
\begin{eqnarray}
\label{eq:growth}
r \frac{d r}{d t} = \frac{D \rho_{v}}{\rho_{l}} (S - 1) \,,
\end{eqnarray}
where $D$ is the molecular diffusion coefficient and $\rho_{v}$ is the vapor density away from the droplet.

\section{Heat transfer from a raindrop}

Let us consider the temperature in a gas layer $T_a(z)$ decreases linearly with increasing height.
\begin{equation}
T_a(z) = T_0 - \beta z \,,
\label{eq:temp}
\end{equation}
The falling droplet will encounter a warming environment and thus will heat up.
The heat transfer rate into the droplet $q$ depends on its mass ($m = \rho_{l} \pi d^3/6$) and its effective heat conductivity $\lambda_{eff} = \lambda \textrm{Nu}$. 
Here, $\textrm{Nu}$ is the Nusselt number, $\lambda$ the thermal conductivity of the ambient gas, $d$ the droplet diameter and $\rho_{l}$ its density. 
We can then write:
\begin{equation}
q = c_p m \frac{dT}{dt} = 2\lambda_{eff} \pi d^2 \frac{T_a-T}{d} \,,
\label{eq:q}
\end{equation}
where, $T$ is the droplet temperature and $c_{p,l}$ specific heat of the liquid. 
We assume that the temperature inside the droplet is homogeneous due to shear enhanced mixing. 
The Nusselt number which expresses the ratio of convective and conductive heat transfer can be estimated for a falling sphere as [21]
\begin{eqnarray}
\label{eq:Nusselt}
\textrm{Nu} = 2 + (0.4 \textrm{Re}^{1/2} + 0.06 \textrm{Re}^{2/3}) \textrm{Pr}^{0.4} \,,
\end{eqnarray}
where $\textrm{Re} = \rho U_t d/\mu$ is the Reynolds number, $\textrm{Pr} = c_{p,g} \mu/\lambda$ is the Prandtl number, and $\rho$, $\mu$, $c_{p,g}$ are the density, the dynamic viscosity, and the specific heat of the ambient gas, respectively, $U_t$ is the terminal velocity of the drop. The drag coefficient required to calculate the terminal velocity iteratively from the equations of motion was obtained from [15]
Using Eq.~(\ref{eq:temp}) and rearranging the terms in Eq.~(\ref{eq:q}), we get:
\begin{equation}
\frac{dT}{dt} = \frac{6 \lambda \textrm{Nu}}{c_{p,l}\rho_{l} d^2} (T_0 - \beta z - T) \,.
\label{eq:dT}
\end{equation}
After the droplet has reached its terminal velocity $U_t$, its vertical position is given by 
\begin{eqnarray}
\label{eq:z_t}
z = z_0 - U_t t \,.
\end{eqnarray}
We thus get
\begin{equation}
\frac{dT}{dt}  = A [T_0-T+\beta(U_t t- z_0)] \,, \,\,
\textrm{with}\,\, A = \frac{6\lambda \textrm{Nu}}{c_{p,l}\rho_{l} d^2} \,.
\label{eq:final}
\end{equation}
Using Eqs.~(\ref{eq:temp}), (\ref{eq:dT}) we can write an evolution equation for the temperature difference between the droplet and local ambient temperature $\Delta = T_a - T$  as
\begin{equation}
\frac{d}{dt}\Delta = \beta U_t - A \Delta \,,
\label{eq:Delta_DEQ}
\end{equation}
resulting in
\begin{equation}
\Delta = \Delta_0e^{-A t} + \frac{\beta U_t}{A} (1 - e^{-A t}) \,,
\label{eq:Delta}
\end{equation}
where $\Delta_0 = \Delta(t=0)$. 
For the temperature profile in atmospheric clouds $\beta$ being the lapse rate, $\beta\approx 5\times 10^{-3}\,^\circ$C/m. 
The material parameters for water and water vapor evaluated at 10$\,^\circ$C are the following:
$\rho_l$ = 10$^3$~kg/m$^{3}$, $c_{p,l}$=4188~kg/m$^3$, $\lambda$ = 0.02 W/(m k).

\section{Melting of an ice particle in a warm cloud}

Within earth's atmospheric conditions an ice particle would start to melt as the temperature increases above 0$\,^\circ$C. The melt water collects around the ice particle. The surface temperature for this ice-water system will not increase until all the ice has melted [24]
. Let us assume that the ice particle has attained its terminal velocity before entering the warm parts of a cloud, then we can state that the amount of energy required to melt the ice particle should balance the net heat transferred into this system from the ambient.
\begin{equation}
\frac{\pi d^3}{6}\rho_{ice} l_{ice} = \int_o^t\frac{Nu \lambda}{d} (T_a - T_{ice})\pi d^2\mathrm{d}t
\label{eq:ice}
\end{equation}
where, $l_{ice}$ represents the specific enthalpy of fusion of ice, $\rho_{ice}$ is the density of the ice. $T_{ice}$ is 0$\,^\circ$C. Using equation~\ref{eq:temp} and integrating equation~\ref{eq:ice} with respect to time we get
\begin{equation}
t = \sqrt[2]{\frac{d^2~\rho_{ice}~\l_{ice}}{3 Nu~\beta~\lambda~U_t}}
\label{eq:ice_time}
\end{equation}
$\rho_{ice}$ = 916 kg m$^{-3}$, l$_{ice}$ = 335 KJ kg$^{-1}$
\end{document}